%% file: carroll_string_160527-arXiv.tex
\documentclass[letterpage,titlepage,11pt]{article}
\usepackage{amsmath,amssymb,amsfonts,color,epsf}
\usepackage{amssymb,slashed,latexsym,multirow,color}
\usepackage{graphics} 
\usepackage{cite}
\usepackage[font=footnotesize,labelsep=newline,labelfont=sc,justification=centering,position=top]{caption}
\input{commands}

\usepackage{color}
\usepackage[
      colorlinks=true,
      linkcolor=blue,
      urlcolor=blue,
      filecolor=black,
      citecolor=green,
      pdfstartview=FitV,
      linktocpage,
      pdftitle={},
        pdfauthor={x, x, x},
        pdfsubject={},
        pdfkeywords={},
        pdfpagemode=None,
        bookmarksopen=true
      ]{hyperref}
      
\makeatletter
\@addtoreset{equation}{section}
\makeatother

\begin{document}

\begin{titlepage}

\begin{flushright}
\hfill{ \hfill{ICCUB-16-018}
\ \ \ \    \ \ \ \ \\
\ \ \ \     \ \ \ }
\end{flushright}
\vskip -1cm

\leftline{}
\vskip 2cm
\centerline{\LARGE \bf Dynamics of Carroll Strings}
\vskip 1.2cm
\centerline{\large\bf Biel Cardona$^{a}$,\ Joaquim Gomis$^{a,b}$ and \ Josep M$^{\text{a}}$ Pons$^{a}$\ \ }
\vskip 0.5cm
\centerline{\sl $^{a}$Departament de F\'{\i}sica Cu\`antica i Astrof\'{\i}sica}
\centerline{\sl and Institut de Ci\`encies del Cosmos,}
\centerline{\sl Universitat de Barcelona, Mart\'i i Franqu\`es 1, E-08028 Barcelona, Spain}
\centerline{\sl $^{b}$Department of Physics, Faculty of Science,}
\centerline{\sl Chulalongkorn University, Bangkok 10330, Thailand}
\smallskip
\vskip 0.5cm

\vskip 1.2cm
\centerline{\large\bf Abstract} \vskip 0.4cm 
\noindent We construct the canonical action of a Carroll string doing the Carroll limit of a canonical 
relativistic string. We also study the Killing symmetries of the Carroll string, which close under an 
infinite dimensional algebra. The tensionless limit and the Carroll $p$-brane action are also discussed.

\end{titlepage}
 
\pagestyle{plain}
\setcounter{page}{1}

\vspace{2mm} \hrule \vspace{1mm} \noindent

\tableofcontents

\vspace{0.7cm}{\hrule \noindent}

\section{Introduction}
The extension of the holographic ideas to non-AdS situations with applications to condensed matter systems 
has produced a renewed interest in the study non-relativistic 
symmetries\footnote{The use of non-relativistic extended objects has been also studied as a soluble 
sector of string theory \cite{Gomis:2000bd}  \cite{Danielsson:2000gi}  \cite{Gomis:2004pw} 
\cite{Gomis:2005pg}  \cite{Brugues:2006nh}.} and the use of non-relativistic gravity theories in the bulk \cite{Son:2013rqa} like Newton-Cartan \cite{Cartan} and Horava gravities \cite{Horava:2009uw}. Recently these theories have been constructed from the gauging of the Bargmann algebra \cite{Andringa:2010it}, from Lihshitz holography \cite{Christensen:2013lma} and from the use non-relativistic conformal methods \cite{Afshar:2015aku}.
 
The study of space-time holography \cite{Banks:2003vp} \cite{deBoer:2003vf} \cite{Arcioni:2003xx} 
\cite{Barnich:2010eb} has lead to reconsider the role of the BMS group \cite{BMS-1}. On the other hand 
it has been shown that BMS symmetry is an infinite conformal extension of the Carroll symmetry 
\cite{Duval:2014}. Carroll symmetry was introduced in \cite{Levy-Leblond} \cite{Bacry:1968zf} as 
the limit of the Poincar\'e algebra when the velocity of light tends to zero. 
There is a duality between the non-relativistic symmetry and the Carroll symmetry \cite{Duval:2014uoa}.  
 
 The strong coupling limit of gravity \cite{Henneaux:1979vn} introduced many years ago 
 was the first example of a dynamical system possessing Carroll symmetry. More recently it has been 
 constructed the action of the Carroll particle \cite{Duval:2014uoa} \cite{old} \cite{Bergshoeff:2014jla} 
 and the Carroll superparticle \cite{Bergshoeff:2015wma}, both exhibiting a trivial dynamics. 
 Notice that in these cases the massless limit can be taken at the level of the action.
 
 The reason for the trivial dynamics of the free particle Carroll objects is due to the fact that the light 
 cone in the Carroll case collapses to the time vertical axis. Carroll symmetries appear also in 
 warped conformal field theories \cite{Hofman:2014loa}. The construction of a Carroll gravity by a modified gauging of the Carroll algebra has been studied recently in \cite{Hartong:2015xda}.
 
 In this note we continue the study of dynamical objects with Carroll symmetry. We construct the action 
 of a tension-full and a tensionless\footnote{The situation regarding the tensionless limit is rather 
 different here than in the case of the non-relativistic limit \cite{Gomis:2004ht}.} 
 Carroll string by taking the Carrollian limits\footnote{Like in the non-relativistic case, where there is not a unique limit for extended objects \cite{Yastremiz:1991jp}  \cite{Gomis:2000bd} \cite{Gomis:2004pw}, the same is true for the Carroll limit of an extended object.} of the canonical action of a relativistic string. We will also construct the action of a Carroll $p$-brane by the same procedure. 
The action for these objects can be also constructed using the method of non-linear realizations \cite{Coleman} applied 
to the Carroll algebras; for the Carroll algebra see \cite{tonnis}, for string Carroll algebras \cite{old}. In either of the two limits, the Carroll string exhibits a trivial dynamics like the Carroll particle. This result also applies to branes. We will also study the Killing symmetries of the Carroll string, 
and show that these symmetries close under an infinite dimensional algebra.

The organization of the paper is as follows, in Section \ref{CCSa} we construct canonical Carroll 
string actions (at least two types of limits form the relativistic string are available). 
In section \ref{KS} we study their Killing symmetries. 
The tensionless limit is analyzed in section \ref{tensionlessCS} and section \ref{Cp} is devoted to 
construct a Carroll $p$-brane action. Finally we write some conclusions and outlook.

\section{Canonical Carroll String action}\label{CCSa}
Our starting point is the canonical action of a relativistic string: \beq\label{cnstring}
S_{\text{NG}} = \int\dd^2\sigma\(p\cdot\dot{x} - \frac {\tilde e}2 \tilde H-\tilde\mu\tilde T\)
= \int\dd\tau\dd\sigma\(p\cdot\dot{x}- \frac{\tilde e}{2}\(p^2 + T^2_r {x'}^2\) - \tilde{\mu}(p\cdot x') \),
\enq
where $\tilde H=p^2 + T^2_r {x'}^2$ and $\tilde T=p\cdot x' $ are the diffeomorphism constraints and ${\tilde{e}}, \tilde{\mu}$ are Lagrange multipliers. In order to obtain the Carroll action for the string we take the `stringy' Carrollian limit by rescaling the longitudinal coordinates $x^\mu$ ($\mu = 0,1$) with a dimensionless parameter $\omega$: \beq\label{stringSet}
{x^\mu} = \frac{X^\mu}{\om}, \hs{0.5} {p_\mu} = \om P_\mu.
\enq
The action is obtained by plugging these expressions in (\ref{cnstring}) and taking the limit 
$\om\riga\infty$. Then the products $p\cdot\dot{x}$ and $p\cdot{x'}$ remain unaffected and become 
$P\cdot\dot{X}$ and $P\cdot{X'}$ respectively. All physics in the ultra-relativistic Carrollian regime 
arises from the constraint proportional to $e$, like in the Carroll particle \cite{Bergshoeff:2014jla}. 
We must rescale the einbein field as in the case of the Carroll particle, $\tilde{e} = e/\om^2$, 
whereas $\tilde{\mu} = \mu$ remains the same. Rescaling the string tension as $T_r = \om T$ and 
sending $\om$ to infinity, we obtain the action of the Carroll string \beq\label{scarroll}
S_{\text{C}_s} = \int\dd\tau\dd\sigma\(P\cdot\dot{X} - \mu(P\cdot X') - 
\frac{e}{2}\(\eta_{\munu}P^\mu P^\nu + T^2{X'_i}^2\)\).
\enq
where ${X'_i}^2 = \delta_{ij}{X^i}'{X^j}'$, $i,j = 2,\dots,D-1$. The transversality constraint does not change, whereas the mass-shell constraint becomes ($E=P^0$): \beq\label{cconstr}
\hat H=-E^2 + (P^1)^2 + T^2{X'_i}^2.
\enq
Notice the absence of the transverse momenta $P_i$.

If instead we perform on the action (\ref{cnstring}) the Carroll limit \emph{\`a la particle} \cite{Bergshoeff:2014jla}, $\displaystyle x^0 = t/\omega$, $p_0=-\omega E$, we get \beq\label{scarroll1}
\tilde S_{\text{C}_s} = \int\dd\tau\dd\sigma\(P\cdot\dot{X} - \mu(P\cdot X') - \frac{e}{2}\(-E^2+ T^2{X'_{\hat\imath}}^2\)\),
\enq
$\hat{\imath} = 1,\dots,D-1$, thus the mass-shell constraint {is now given by} $ \tilde H=-E^2+ T^2{X'_{\hat\imath}}^2$. 

In the Carroll case these two limits give actions with 
the same physics as regards the equations of motion. This is in sharp contrast with the non-relativistic 
case, where the 
limits \emph{\`a la string} or \emph{\`a la particle} lead to different dynamics: in one case that of the 
vibrating string ((\emph{`stringy'} NR limit) \cite{Gomis:2000bd} \cite{Gomis:2005pg} \cite{Gomis:2004pw}, 
and in the other a non-vibrating string with a fixed length ({\emph{\`a la particle} limit) 
\cite{Yastremiz:1991jp} \cite{carlesdaniel}. We observe nevertheless that a physical difference 
between the two Carroll strings appears when we dimensionally reduce 
on $S^1$: in one case we obtain a Carroll massless particlewhereas in the other a massive 
Carroll particle \cite{Bergshoeff:2014jla} \cite{Duval:2014}.

\subsection{Carroll Symmetries, Carroll String algebra and Carroll Diffeomorphism}
The canonical action (\ref{scarroll}) is invariant under the \emph{`stringy'} Carroll 
transformations: \beq\begin{split}\label{carrolltrans}
\delta X^\mu  &= \om^\mu{}_{\nu}X^\nu+  \om^\mu{}_{i}X^i+ \zeta^\mu, \hs{0.75} \delta X^i = \om^i_{\ j}X^j + \zeta^i, \\
\delta P_\mu &= \om_\mu^{\ \nu}P_\nu, \hs{3.4} \delta P_i = \om_i{}^{ \mu}P_\mu +\om_i{}^{j}P_j,
\end{split}\enq
where $( \om^\mu{}_{\nu}, \om^\mu{}_{i}, \om^i_{\ j}, \zeta^\mu, \zeta^i)$ are respectively the 
Lorentz boosts in the two longitudinal directions, the time and space Carroll boosts, 
the spatial rotations, longitudinal translations and the transverse translations. These transformations 
can all be derived from a general infinitesimal Poincar\'e transformation, $\delta x^M = \om^M_{ \ N}x^N + \xi^M$ and $\delta p_M = \om_M^{ \ N}p_N$, by performing the rescaling \beq
x^\mu = \frac{X^\mu}{\lambda}, \hs{0.75} \om^\mu_{\ i} \riga \frac{\om^\mu_{\ i}}{\lambda}, \hs{0.75} \xi^\mu = \frac{\zeta^\mu}{\lambda}, \hs{0.75} p_\mu = \lambda P_\mu,
\enq
and taking the limit $\lambda\riga\infty$. 

The algebra of these transformations {closes} under what we call the String Carroll algebra \beq\begin{split}\label{stringcarroll}
[M_{ij},P_k] &= 2\delta_{k[j}P_{i]}, \hs{0.75} \[M_{ij},M_{kl}\]Ê= 2\delta_{i[k}M_{l]j} - 2\delta_{j[k}M_{l]i}, \\
[K,P_0] &= P_1, \hs{2.2} [K,P_1] = P_0, \\
[K,K_i] &= B_i, \hs{2.21} [K, B_i] = K_i, \\
[M_{ij},K_k] &= 2\delta_{k[j}K_{i]}, \hs{0.95} [M_{ij},B_k] = 2\delta_{k[j}B_{i]}, \\
[K_i,P_j] &= -\delta_{ij}P_0, \hs{1.42} [B_i,P_j] = -\delta_{ij}P_1 \hs{0.25}
\end{split}\enq
where Lie algebra generators are the longitudinal Lorentz boost $K$, the time Carroll boosts $K_i$, 
the space Carroll boosts $B_i$, the spatial rotations $M_{ij}$, and the time-space translations $P_0, P_1, P_i$. This algebra can be obtained from the Poincar\'e algebra \beq\begin{split}\label{poincalg}
\[M_{AB},M_{CD}\] &= 2\eta_{A[C}M_{D]B} - 2\eta_{B[C}M_{D]A}, \\
\[M_{AB},P_C\] &= 2\eta_{C[B}P_{A]}, \hs{0.5} \[P_A,P_B\] = 0,
\end{split}\enq
by the contraction \beq
P_A \riga (\om P_0,\om P_1,P_i), \hs{0.5} K_i \riga \om K_i, \hs{0.5} B_i \riga \om B_i
\enq
with the identifications $M_{01} \equiv K$, $M_{i0} \equiv K_i$, $M_{i1} \equiv B_i $.

The action has also the gauge invariance of diffeomorphisms. These are generated by the first 
class constraints $\hat H(\sigma)=-E^2 + (P^1)^2 + T^2{X'_i}^2,\, \hat T(\sigma)=(P\cdot X')$, 
whose algebra is \beq\begin{split}\label{diffalgebra}
\{\hat{H}(\sigma),\hat{H}(\sigma')\} &= 0, \\
\{\hat{H}(\sigma),\hat{T}(\sigma')\} & = \hat{H}(\sigma)\partial_\sigma\delta (\sigma-\sigma') - \hat{H}(\sigma')\partial_{\sigma'}\delta (\sigma-\sigma'), \\
\{\hat{T}(\sigma),\hat{T}(\sigma')\} & = \hat{T}(\sigma)\partial_\sigma\delta (\sigma-\sigma') - \hat{T}(\sigma')\partial_{\sigma'}\delta (\sigma-\sigma').
\end{split}\enq

\subsection{Dynamics}\label{eom}
The action of the Carroll string is given by (\ref{scarroll}). The canonical Hamiltonian is: \beq
H_D = \int\dd\sigma\(\mu\(\eta_{\munu}P^\mu{X^\nu}' + P_i{X^i}'\) + \frac{e}{2}\(\eta_{\munu}P^\mu P^\nu + T^2{X'_i}^2\)\).
\enq
The non-vanishing Poisson brackets are given by \beq\begin{split}\label{PBxp}
\{X^M(\sigma),P^N(\sigma')\} &= \eta^{MN}\delta(\sigma-\sigma'), \\
\{X^M(\sigma),X^N(\sigma')\} &= \{P_M(\sigma),P_N(\sigma')\} = 0, \\
\end{split}\enq
the equations of motion follow: 
\beq\begin{split}\label{eomcf}
\dot{X}^\mu &= \mu {X^\mu}' + eP^\mu, \hs{0.75} \dot{P}^\mu = (\mu P^\mu)', \\
\dot{X}^i &= \mu {X^i}', \hs{2.05} \dot{P}^i = \(\mu P^i + eT^2{X^i}'\)', \\
\end{split}\enq
and the constraints \beq
\eta_{\munu}P^\mu{X^\nu}' + P_i{X^i}'=0, \hs{0.5} \eta_{\munu}P^\mu P^\nu + T^2{X'_i}^2=0.
\enq

As we can see the dynamics of the Carroll string is trivial. In fact considering the analogous of 
the conformal gauge, $e=1,\ \mu=0$, we have \beq\begin{split}\label{eomconformal}
\dot{X}^\mu &= P^\mu, \hs{0.75} \dot{P}^\mu = 0, \\
\dot{X}^i &= 0, \hs{1.13} \dot{P}^i = T^2\({X^i}'\)'.
\end{split}\enq
We see that the transverse coordinates of the bosonic string are constant, therefore the free Carroll 
string does not move. Notice however that the momenta are not constant. This is a common feature of (free) 
Carroll particle, Carroll string or in general, Carroll $p$-brane (see section \ref{Cp}): in Carroll space there is no connection between spatial momenta and velocities.
 
If we consider Carroll strings coupled to Carroll gravity the strings will have a non-trivial dynamics 
like in the case of the Carroll particle coupled to Carroll gauge fields \cite{Bergshoeff:2014jla}.
 
\section{Killing Symmetries of the Carroll String}\label{KS}
In this section we analyze the Killing symmetries of the Carroll string. As we will see the string action 
is invariant under an infinite dimensional group of transformations that includes the String Carroll 
transformations (\ref{carrolltrans}). It turns out that the full symmetry group includes conformal 
symmetries in both the transverse and the longitudinal fields. In contrast to higher dimensions, 
the conformal algebra in two dimensions is infinite dimensional, hence longitudinal fields will have 
a infinite-dimensional symmetry, a common feature in the Carrollian context \cite{Bergshoeff:2014jla} \cite{Bergshoeff:2015wma}. This is not the case for the transversal $D-2$ coordinates.

Let us consider the generator of canonical symmetry transformations \beq\label{generator}
G = \int\dd\sigma\(\xi^MP_M + \La\pi_e + \ga\pi_\mu\),
\enq
with $\xi^M$, $\Lambda$ and $\gamma$ arbitrary functions on the extended configuration space, which
includes $e$ and $\mu$ as new variables. 
The extended phase space includes the momenta $\pi_e,\ \pi_\mu$, which are the new primary constraints. 
Conservation of $G$ reads: 
\beq\begin{split}\label{killstring}
0 &= \dot{G} = \int\dd\sigma\bigg(eP^\mu P^\nu\(2\partial_{(\mu}\xi_{\nu)} - \frac{\Lambda}{2e}\eta_{\munu}\) + e\(P^iP^\mu - T^2{X^\mu}'{X^i}'\)\partial_\mu\xi_i \\
&\hs{2.25}- eT^2{X^i}'{X^j}'\(2\partial_{(i}\xi_{j)} + \frac{\Lambda}{2e}\delta_{ij}\) -\gamma\(\eta_{\munu}P^\mu{X^\nu}' + P_i{X^i}'\)\bigg).
\end{split}\enq
The Killing equations are: 
\beq\label{killing}
\partial_\mu\xi_\nu +\partial_\nu\xi_\mu  = \tilde{\lambda}\eta_{\munu}, \hs{0.75} \partial_\mu\xi_i = 0, \hs{0.75} \partial_i\xi_j + \partial_j\xi_i = - \tilde{\lambda}\delta_{ij}, \hs{0.75} \gamma = 0,
\enq
with $\Lambda$ determined as $\Lambda=e\tilde{\lambda}$ and with the conformal factor $\tilde{\lambda}$ (notice that $\tilde{\lambda} = \tilde{\lambda}(X^i)$) satisfying: \beq
\tilde{\lambda} = \partial_\mu\xi^\mu = -\frac{2}{D-2}\partial_i\xi^i.
\enq

The second Killing equation tells us that $\xi^i = \xi^i(X^j)$. Notice also the sign difference
in (\ref{killing})
between the conformal Killing equation for longitudinal vectors and transversal ones. In particular when we consider a scale transformation we have Lifshitz scaling with $z=-1$: $\delta X^\mu=-c X^\mu$ and $\delta X^i= c X^i$, where $c$ is the infinitesimal dilatation parameter. The equation for transverse fields is the Euclidean Conformal Killing equation. For $D = 3$ there is only one transversal direction and hence no restriction on $\xi^i$. For $D=4$ we have the standard two-dimensional infinite conformal symmetry for the (two) transverse variables. For $D > 4$, we get \beq
\tilde{\lambda}(X^i) = 2\big(2b_kX^k - c\big)\,,
\enq
for some constants $b_k$ and $c$. In this case the solution is: \beq\label{soltransv}
\xi^i(X^k) = a^i + \om^i_{\ j}X^j + cX^i + X^2_j b^i - 2(b_jX^j)X^i,
\enq
where $a^i$, $\om^i_{\ j}$, $c$ and $b^i$ generate space translations, rotations, space dilatations and special conformal transformations, respectively. Notice as a feature of the rescaling $\om\riga\infty$ of the relativistic string 
to obtain the Carroll string (as done in section \ref{CCSa}) that if we dimensionally reduce the Carroll string to the Carroll particle, this reduction at the level of the Killing symmetries does not reproduce the infinite-dimensional symmetry for transverse fields which exists in the particle case \cite{Bergshoeff:2014jla}.

The solution to the first equation is: \beq
\xi^\mu(X) = \frac{\tilde{\lambda}}{2}X^\mu + \Omega^\mu_{\ \nu}(X^i)X^\nu + f^\mu(X^i),
\enq
where the antisymmetric tensor $\Omega_{\mu \nu}$ and the vector $f^\mu$ have arbitrary dependences on the transverse coordinates $X^i$.

In case we consider the action (\ref{scarroll1}) for the Carroll string \emph{\`a la particle}, the Killing equations give the following transformations \beq
\xi^0(X) = \frac{\tilde{\lambda}}{2}X^0 + f^0(X^i),
\enq
and the same results as before for the spatial components $\xi^i(X)$, but now for $i=1,2,\dots,D-1$.  

\section{Tensionless Carroll String}\label{tensionlessCS}
Like in the particle case where the massless limit can be taken straightforwardly, here we analyze 
the tensionless limit\footnote{The tensionless limit of the relativisic string and branes has
been widely dicussed in the literature, 
see for example  \cite{Gamboa:1989zd}\cite{Bandos:1993ma}\cite{Zheltukhin:1997wj}.}, $T\riga 0$, of the Carroll string action (\ref{scarroll}). 
In this limit the mass-shell becomes $\h{H}_0 = -(P^0)^2 + (P^1)^2$ and the Dirac's Hamiltonian is: \beq
H_D(T\riga0) = \int\dd\sigma\(\mu\(\eta_{\munu}P^\mu{X^\nu}' + P_i{X^i}'\) + \frac{e}{2}\eta_{\munu}P^\mu P^\nu\).
\enq
The equations of motion are: \beq\begin{split}\label{eomcf-less}
\dot{X}^\mu &= \mu {X^\mu}' + eP^\mu, \hs{0.75} \dot{P}^\mu = (\mu P^\mu)', \\
\dot{X}^i &= \mu {X^i}', \hs{2.06} \dot{P}^i = \(\mu P^i\)'.
\end{split}\enq
Again, the dynamics is trivial. Taking the conformal gauge we see that the string does not move. 
Additionally in this case the momenta are also constant.

Let us study the Killing symmetries of this system.  Considering the same generator of symmetry transformations as before (\ref{generator}), conservation of $G$ leads to: 
\beq\begin{split}\label{killstring-less}
0 &= \dot{G} = \int\dd\sigma\bigg(eP^\mu P^\nu\(\partial_{(\mu}\xi_{\nu)} - 
\frac{\Lambda}{2e}\eta_{\munu}\) + eP^iP^\mu\partial_\mu\xi_i -\gamma\(\eta_{\munu}P^\mu{X^\nu}' 
+ P_i{X^i}'\)\bigg),
\end{split}\enq
and the Killing equations are: 
\beq\label{killing-less}
\partial_\mu\xi_\nu +\partial_\nu\xi_\mu  = 
\tilde{\lambda}\eta_{\munu}, \hs{0.75} \partial_\mu\xi_i = 0, \hs{0.75} \gamma = 0,
\enq
with conformal factor $\tilde{\lambda} = \Lambda/e = \partial_\mu\xi^\mu$. 
Thus we obtain the standard two dimensional conformal symmetry -as expected because the 
dimensionful parameter $T$ has been eliminated- where the ``holomorphic'' and ``anti-holomorphic'' functions have an arbitrary dependence on the transverse coordinates. On the other hand, the functions $\xi_i(X^j)$ are arbitrary.

\section{ Carroll $p$-brane Action}\label{Cp}
The construction of Carroll $p$-branes follows the same steps of the string case. The canonical $p$-brane 
action in $D$-dimensional ($D>p$) Minkowski space is: \beq\begin{split}\label{p-action}
S_{p\text{-brane}} &= \int\dd^{p+1}\xi\(p\cdot\dot{x} - \tilde{s}^{\bar{a}}\HH_{\bar{a}} - \frac{\tilde{v}}{2}\HH\) \\
&= \int\dd\tau\dd^p\sigma\(p\cdot\dot{x} - \tilde{s}^{\bar{a}}\(p\cdot\partial_{\bar{a}} x\) - 
\frac{\tilde{v}}{2}\Big(p^2 + T^2_p\det(g_{\bar{a}\bar{b}})\Big)\),
\end{split}\enq
where $\HH=p^2 + T^2_p\det(g_{\bar{a}\bar{b}})$ and $\HH_{\bar{a}}=p\cdot\partial_{\bar{a}} x$ are the diffeomorphism constraints and $\tilde{s}^{\bar{a}}$ and $\tilde{v}$ are the $p+1$ Lagrange multipliers. The metric $g_{\bar{a}\bar{b}} = \partial_{\bar{a}} x^M\partial_{\bar{b}}x^N \eta_{MN}$ ($\bar{a},\bar{b} = 1,\dots,p$ is the induced metric on the worldspace. Now we consider the Carrollian $p$-brane limit, \beq\label{pbranelimit}
x^\mu = \frac{X^\mu}{\om}, \hs{0.75} p_\mu = \om P_\mu.
\enq
The quantities $p\cdot\dot{x}$ and $p\cdot\partial_{\bar{a}}x$ do not change and
become $P\cdot\dot{X}$ and $P\cdot\partial_{\bar{a}}X$ respectively. But the last constraint in 
(\ref{p-action}) changes. The rescaling on the Lagrange multipliers 
$\tilde{s}^{\bar{a}}$, $\tilde{v}$ as well as the $p$-brane tension are the same as in the string case. We have: $\tilde{s}^{\bar{a}} = s^{\bar{a}}$, $\tilde{v} = v/\om^2$ and $T_p = \om T$. In the limit $\om\riga\infty$ we obtain ($i,j = p+1,\dots,D-1$): \beq
\frac{\tilde{v}}{2}\(p^2 + T_p^2\det(g_{\bar{a}\bar{b}})\) = \frac{v}{2}\(\eta_{\munu}P^\mu P^\nu 
+ T^2\,\gamma\),
\enq
with $\gamma = \det(\gamma_{\bar a \bar b}) = \det(\partial_{\bar{a}} X^i\partial_{\bar{b}} X^j \delta_{ij})$. The Carroll $p$-brane action turns out to be: \beq
S_{\text{C}_p} = \int\dd\tau\dd^p\sigma\(P\cdot\dot{X} - s^{\bar{a}}\(P\cdot\partial_{\bar{a}} X\) 
- \frac{v}{2}\(\eta_{\munu}P^\mu P^\nu + T^2\gamma\)\).
\enq

At this point, we do not need to do a full analysis. Notice that the behaviour of the tension in the Carroll limit does not depend on $p$, the number of dimensions of the worldspace. It is the same rescaling for the point particle (in this case $T = M$)  \cite{Bergshoeff:2014jla} \cite{Bergshoeff:2015wma}, for the string, and so on. The substantial difference between the particle and the string is in the rescaling of $X^1$. If we rescale the first $p$ spatial coordinates, we can expect that the derived action will contain the same physics as that for the string. The results in the section above also hold if we add $p$ spatial extra-dimensions. 
This behaviour differs from the non-relativistic case, 
where the rescaled $p$-brane tension is $T_p = \om^{1-p}T$ \cite{Gomis:2004pw}. 
Notice that the behaviour of the tension in the Carroll limit does not depend on $p$, the number 
of dimensions of the worldspace. It is the same rescaling for the point particle (in this case $T = M$)  
\cite{Bergshoeff:2014jla} \cite{Bergshoeff:2015wma}, for the string, and so on.

The dynamics of the Carroll $p$-brane, like the Carroll particle or the Carroll string, is also 
trivial. In the conformal gauge, 
$ v=1,\ s^{\bar{a}}=0$, the equations of motion are
\beq\begin{split}\label{eomconformal-p}
\dot{X}^\mu &= P^\mu, \hs{0.75} \dot{P}^\mu = 0, \\
\dot{X}^i &= 0, \hs{1.15} \dot{P}^i = {\partial_{\bar a}\left(v\, T^2 \gamma\, \gamma^{\bar a \bar b}\partial_{\bar{b}} X^i\right)}.
\end{split}\enq 

\section{Discussion and Outlook}\label{DO}
We have constructed the action of  tension-full and tensionless Carroll extended objects by doing the 
different Carrollian limits of a relativistic string or a $p$-brane canonical action. 
The action for the tension-full objects can be also constructed using the method of non-linear 
realizations applied to the Carroll algebras, for the Carroll algebra see \cite{tonnis}, for string 
Carroll algebras \cite{old}. The dynamics of the ($p$-brane) string Carroll 
actions are trivial independently if one considers the Carroll limit \emph{\`a la particle} 
\cite{Levy-Leblond} or the ($p$-brane) \emph{'stringy'} Carroll limit, 
(\ref{pbranelimit}) (\ref{stringSet}). The reason for the trivial dynamics 
for these free Carroll dynamical objects is due to the fact that the light cone in the Carroll 
case collapses to the time vertical axis. In contrast, in the non-relativistic case the string and particle limit lead to different non-relativistic models, one being the vibrating string (with the string NR limit) \cite{Gomis:2000bd} \cite{Gomis:2005pg} \cite{Gomis:2004pw} and the other a non-vibrating string with a fixed lenght (particle limit) \cite{Yastremiz:1991jp} \cite{carlesdaniel}.

If we consider the coupling of the Carroll extended objects to Carroll gauge fields as it is done for the case the particle in \cite{Bergshoeff:2014jla}, the dynamics becomes non-trivial \cite{eric+col} because the interaction with the Carroll background fields opens the light cone. Extended objects with a Carroll supersymmetry that generalize the Carroll superparticle \cite{Bergshoeff:2015wma} can also be studied.

\section*{Acknowledgements}\addcontentsline{toc}{section}{Acknowledgements}
We acknowledge discussions with Eric Bergshoeff, Carles Batlle, Blaise Rollier, Jan Rosseel, Tonnis Ter Veldhuis. JG acknowledges Auttakit Chatrabhuti  and Oleg Evin for the hospitality at the Department of Physics of Chulalongkorn University and for discussions. We acknowledge financial support from projects FPA2013-46570, 2014-SGR-104 and MDM-2014-0369 of ICCUB (Unidad de Excelencia ÔMaria de MaeztuÕ). JG also acknowledges financial support from CUniverse research promotion project by Chulalongkorn University (grant reference CUAASC).


\end{document}

%% file: commands.tex
\newcommand{\munu}{\mu\nu}

\newcommand{\h}{\hat}

\newcommand{\om}{\omega}		\newcommand{\ga}{\gamma}
\newcommand{\La}{\Lambda}

\newcommand{\dd}{\mathrm{d}} 

\newcommand{\HH}{\mathcal{H}}		
		
\newcommand{\hs}[1]{\hspace{#1 cm}}		

\renewcommand{\(}{\left(}			\renewcommand{\)}{\right)}
			\renewcommand{\]}{\right]}
		
\newcommand{\riga}{\rightarrow}		

\newcommand{\beq}{\begin{equation}}		\newcommand{\enq}{\end{equation}}

%% file: carroll_string_160527-arXiv.bbl
\begin{thebibliography}{99}\addcontentsline{toc}{section}{References}
\bibitem{Gomis:2000bd}
 J. Gomis and H. Ooguri, ``Non-relativistic closed string theory,'' {\em J.\ Math.\ Phys.} {\bf 42} (2001) 3127, \href{http://arxiv.org/abs/hep-th/0009181}{{\tt arXiv:0009181 [hep-th]}}.

\bibitem{Danielsson:2000gi}
  U.~H.~Danielsson, A.~Guijosa and M.~Kruczenski,
  ``IIA/B, wound and wrapped,''
  JHEP {\bf 0010} (2000) 020,
  doi:10.1088/1126-6708/2000/10/020,
  \href{http://arxiv.org/abs/hep-th/0009182}{{\tt arXiv:0009182 [hep-th]}}.

\bibitem{Gomis:2004pw}
J.~Gomis, K.~Kamimura and P.~K.~Townsend, ``Non-relativistic superbranes,'' {\em JHEP} {\bf 0411} (2004) 051, \href{http://arxiv.org/abs/hep-th/0409219}{{\tt arXiv:0409.219 [hep-th]}}.

\bibitem{Gomis:2005pg}
 J.~Gomis, J.~Gomis and K.~Kamimura, ``Non-relativistic superstrings: A New soluble sector of AdS$_5 \times S^5$,'' {\em JHEP} {\bf 0512} (2005) 024, \href{http://arxiv.org/abs/hep-th/0507036}{{\tt arXiv:0507036 [hep-th]}}.
  
\bibitem{Brugues:2006nh}
  J. Brugu\'es, J. Gomis, and K. Kamimura, ``Newton-Hooke Algebras, Non-relativistic Branes and Generalized pp-wave Metrics,'' {\em Phys.\ Rev.\ B} {\bf 73} (2006) 085011, \href{http://arxiv.org/abs/hep-th/0603023}{{\tt arXiv:0603023 [hep-th]}}.

\bibitem{Son:2013rqa}
  D.~T.~Son,``Newton-Cartan Geometry and the Quantum Hall Effect,''
  \href{https://arxiv.org/abs/1306.0638}{{\tt arXiv:1306.0638 [cond-mat.mes-hall]}}.

\bibitem{Cartan}
E.~Cartan, {\it {Sur les vari\'et\'es \`a connexion affine et la th\'eorie de la relativit\'e g\'en\'eralis\'ee. (premi\`ere partie)}}, {Annales Sci. Ecole Norm. Sup.} {\bf 40} (1923) 325--412.

\bibitem{Horava:2009uw}
  P.~Horava,
  ``Quantum Gravity at a Lifshitz Point,''
  Phys.\ Rev.\ D {\bf 79} (2009) 084008,
  doi:10.1103/PhysRevD.79.084008,
  \href{https://arxiv.org/abs/0901.3775}{{\tt arXiv:0901.3775 [hep-th]}}.

\bibitem{Andringa:2010it}
  R.~Andringa, E.~Bergshoeff, S.~Panda and M.~de Roo,
  ``Newtonian Gravity and the Bargmann Algebra,''
  Class.\ Quant.\ Grav.\  {\bf 28} (2011) 105011,
  doi:10.1088/0264-9381/28/10/105011,
  \href{http://arxiv.org/abs/1011.1145}{{\tt arXiv:1011.1145 [hep-th]}}.

\bibitem{Christensen:2013lma}
  M.~H.~Christensen, J.~Hartong, N.~A.~Obers and B.~Rollier,
  ``Torsional Newton-Cartan Geometry and Lifshitz Holography,''
  Phys.\ Rev.\ D {\bf 89} (2014) 061901,
  doi:10.1103/PhysRevD.89.061901,
  \href{http://arxiv.org/abs/1311.4794}{{\tt arXiv:1311.4794 [hep-th]}}.
  
\bibitem{Afshar:2015aku}
  H.~R.~Afshar, E.~A.~Bergshoeff, A.~Mehra, P.~Parekh and B.~Rollier,
  ``A Schr\"odinger approach to Newton-Cartan and Ho\v{r}ava-Lifshitz gravities,''
  \href{http://arxiv.org/abs/1512.06277}{{\tt arXiv:1512.06277 [hep-th]}}.
  
\bibitem{Banks:2003vp}
  T.~Banks,
 ``A Critique of pure string theory: Heterodox opinions of diverse dimensions,''
  \href{http://arxiv.org/abs/hep-th/0306074}{{\tt arXiv:0306074 [hep-th]}}.
    
\bibitem{deBoer:2003vf}
  J.~de Boer and S.~N.~Solodukhin,
  ``A Holographic reduction of Minkowski space-time,''
  Nucl.\ Phys.\ B {\bf 665} (2003) 545,
  \href{http://arxiv.org/abs/hep-th/0303006}{{\tt arXiv:0303006 [hep-th]}}.

\bibitem{Arcioni:2003xx}
  G.~Arcioni and C.~Dappiaggi,
  ``Exploring the holographic principle in asymptotically flat space-times via the BMS group,''
  Nucl.\ Phys.\ B {\bf 674} (2003) 553,
  \href{http://arxiv.org/abs/hep-th/0306142}{{\tt arXiv:0306142 [hep-th]}}.
  
\bibitem{Barnich:2010eb}
  G.~Barnich and C.~Troessaert,
  ``Aspects of the BMS/CFT correspondence,''
  JHEP {\bf 1005} (2010) 062,
  \href{http://arxiv.org/abs/1001.1541}{{\tt arXiv:1001.1541 [hep-th]}}.

\bibitem{BMS-1} Bondi H., van der Burg M.G.J., Metzner A.W.K., 1962 Proc. Roy. Soc. A, 269, 21 (paper VII); Sachs R., Proc. Roy. Soc. Lond. {\bf 270}, (1962), 103; Sachs R., Phys. Rev., {\bf 128} (1962), 2851.

\bibitem{Duval:2014}
C.~Duval, G.~W.~Gibbons and P.~A.~Horvathy, ``Conformal Carroll groups,'' \href{http://arxiv.org/abs/1403.4213}{{\tt arXiv:1403.4213 [hep-th]}}; C.~Duval, G.~W.~Gibbons and P.~A.~Horvathy, ``Conformal Carroll groups and BMS symmetry,'' \href{http://arxiv.org/abs/1402.5894}{{\tt arXiv:1402.5894 [gr-qc]}}.

\bibitem{Levy-Leblond}
J.M.~L\'evy-Leblond, ``Une nouvelle limite non-relativiste du group de Poincar\'e'', {\em Ann.~Inst.~H.~Poincar\'e} {\bf 3} (1965) 1; V. D. Sen Gupta, ``On an Analogue of the Galileo Group,'' {\em Il Nuovo Cimento} {\bf 54} (1966) 512.
 
\bibitem{Bacry:1968zf}
H.~Bacry and J.~Levy-Leblond, ``Possible kinematics,'' {\em J.\ Math.\ Phys.} {\bf 9} (1968) 1605.

\bibitem{Duval:2014uoa}
  C.~Duval, G.~W.~Gibbons, P.~A.~Horvathy and P.~M.~Zhang,
  ``Carroll versus Newton and Galilei: two dual non-Einsteinian concepts of time,''
  Class.\ Quant.\ Grav.\  {\bf 31} (2014) 085016,
  doi:10.1088/0264-9381/31/8/085016,
  \href{https://arxiv.org/abs/1402.0657}{{\tt arXiv:1402.0657 [gr-qc]}}.
  
\bibitem{Henneaux:1979vn}
  M.~Henneaux,
  ``Geometry of Zero Signature Space-times,''
  Bull.\ Soc.\ Math.\ Belg.\  {\bf 31} (1979) 47.

\bibitem{old}
J. Gomis and F. Passerini, unpublished notes (2005).

\bibitem{Bergshoeff:2014jla}
E.~Bergshoeff, J.~Gomis and G.~Longhi, ``Dynamics of Carroll Particles,'' {\em Class.\ Quant.\ Grav.} {\bf 31} (2014) 20, \href{http://arxiv.org/abs/1405.2264}{{\tt arXiv:1405.2264 [hep-th]}}.

\bibitem{Bergshoeff:2015wma}
E.~Bergshoeff, J.~Gomis and L.~Parra, ``The Symmetries of the Carroll Superparticle,'' \href{http://arxiv.org/abs/1503.06083}{{\tt arXiv:1503.06083 [hep-th]}}.

\bibitem{Gamboa:1989zd}
  J.~Gamboa, C.~Ramirez and M.~Ruiz-Altaba,
  ``Field Theory of Null Strings and $P^-$branes,''
  Phys.\ Lett.\ B {\bf 231} (1989) 57.
  doi:10.1016/0370-2693(89)90113-5

\bibitem{Bandos:1993ma}
  I.~A.~Bandos and A.~A.~Zheltukhin,
  ``Null super p-branes quantum theory in four-dimensional space-time,''
  Fortsch.\ Phys.\  {\bf 41} (1993) 619.
  doi:10.1002/prop.19930410703
  
  
\bibitem{Zheltukhin:1997wj}
  A.~A.~Zheltukhin,
  ``A Hamiltonian of null strings: An invariant action of null (super)membranes,''
  Sov.\ J.\ Nucl.\ Phys.\  {\bf 48} (1988) 375
   [Yad.\ Fiz.\  {\bf 48} (1988) 587].
  
  
\bibitem{Hofman:2014loa}
  D.~M.~Hofman and B.~Rollier,
  ``Warped Conformal Field Theory as Lower Spin Gravity,''
  Nucl.\ Phys.\ B {\bf 897} (2015) 1,
  doi:10.1016/j.nuclphysb.2015.05.011,
  \href{https://arxiv.org/abs/1411.0672}{{\tt arXiv:1411.0672 [hep-th]}}.
  
\bibitem{Hartong:2015xda}
  J.~Hartong,
  ``Gauging the Carroll Algebra and Ultra-Relativistic Gravity,''
  JHEP {\bf 1508} (2015) 069,
  doi:10.1007/JHEP08(2015)069,
  \href{http://arxiv.org/abs/1505.05011}{{\tt arXiv:1505.05011 [hep-th]}}.
  
\bibitem{Gomis:2004ht}
  J.~Gomis and F.~Passerini,
  ``Rotating solutions of non-relativistic string theory,''
  Phys.\ Lett.\ B {\bf 617} (2005) 182,
  doi:10.1016/j.physletb.2005.04.061,
  \href{http://arxiv.org/abs/hep-th/0411195}{{\tt arXiv:0411195 [hep-th]}}.

\bibitem{Coleman}
  S.~R.~Coleman, J.~Wess and B.~Zumino,
  ``Structure of phenomenological Lagrangians. 1',
  Phys.\ Rev.\  {\bf 177} (1969) 2239;
  \newline
  C.~G.~.~Callan, S.~R.~Coleman, J.~Wess and B.~Zumino,
  {``Structure of phenomenological Lagrangians. 2''},
  Phys.\ Rev.\  {\bf 177} (1969) 2247.


\bibitem{tonnis} 
T.E.~Clark and  T.~ Ter Veldhuis, ``AdS Carroll Branes'' arXiv:1605.05484.

\bibitem{Yastremiz:1991jp}
  C.~F.~Yastremiz,
  ``Galilean extended objects,''
  Class.\ Quant.\ Grav.\  {\bf 9} (1992) 2395,
  doi:10.1088/0264-9381/9/11/007.

\bibitem{carlesdaniel}
C. Batlle, J. Gomis and D. Not,  work in progress.

\bibitem{eric+col} Work in progress.

\end{thebibliography}
